\documentclass[aip,jap,reprint,amsmath,amssymb,floatfix,superscriptaddress]{revtex4-2}

\usepackage[english]{babel}
\usepackage[utf8]{inputenc}
\usepackage[colorlinks,citecolor=blue,linkcolor=blue]{hyperref}
\usepackage{graphicx}	
\usepackage{dcolumn}	
\usepackage{bm}			
\usepackage{txfonts} 	
\usepackage{siunitx}    
\usepackage{color,soul}
\usepackage{xcolor}
\usepackage[pagewise]{lineno}
\usepackage[version=4]{mhchem} 

\newcommand{\MRL}{Materials Research Laboratory, University of Illinois at Urbana-Champaign, Urbana, Illinois 61801, USA}
\newcommand{\MSE}{Department of Materials Science and Engineering, University of Illinois Urbana-Champaign, Urbana, Illinois 61801, USA}
\newcommand{\PHY}{Department of Physics, University of Illinois Urbana-Champaign, Urbana, Illinois 61801, USA}

\begin{document}

\title{Thermal contribution to current-driven antiferromagnetic-order switching}

\author{Myoung-Woo Yoo}
\email[Corresponding author: ]{mwyoo@illinois.edu}
\affiliation{\MRL}
\affiliation{\MSE}

\author{Virginia O. Lorenz}
\affiliation{\MRL}
\affiliation{\PHY}

\author{Axel Hoffmann}
\affiliation{\MRL}
\affiliation{\MSE}
\affiliation{\PHY}

\author{David G. Cahill}
\affiliation{\MRL}
\affiliation{\MSE}
\affiliation{\PHY}

\date{\today}

\begin{abstract}
In information technology devices, current-driven state switching is crucial in various disciplines including spintronics, where the contribution of heating to the switching mechanism plays an inevitable role. Recently, current-driven antiferromagnetic order switching has attracted considerable attention due to its implications for next-generation spintronic devices. Although the switching mechanisms can be explained by spin dynamics induced by spin torques, some reports have claimed that demagnetization above the N{\'e}el temperature due to Joule heating is critical for switching. Here we present a systematic method and an analytical model to quantify the thermal contribution due to Joule heating in micro-electronic devices, focusing on current-driven octupole switching in the non-collinear antiferromagnet, {\ce{Mn3Sn}}. The results consistently show that the critical temperature for switching remains relatively constant above the N{\'e}el temperature, while the threshold current density depends on the choice of substrate and the base temperature. In addition, we provide an analytical model to calculate the Joule-heating temperature which quantitatively explains our experimental results. From numerical calculations, we illustrate the reconfiguration of magnetic orders during cooling from a demagnetized state of polycrystalline {\ce{Mn3Sn}}. This work not only provides deeper insights into magnetization switching in antiferromagnets, but also a general guideline for evaluating the Joule-heating temperature excursions in micro-electronic devices.

\end{abstract}

\maketitle
\section{Introduction}
Novel non-volatile data storage technologies, including magnetic random-access memory, resistive random-access memory, and phase-change memory, are fundamentally based on current-induced resistance changes~\cite{bhatti_spintronics_2017, zahoor_resistive_2020, le_gallo_overview_2020}. Despite the different switching mechanisms inherent in each technology, the contribution of heating to the mechanisms is inevitable and has been a central focus of research in various disciplines~\cite{zimmers_role_2013, li_roles_2018, zhang_effect_2020, pham_thermal_2018}. With the growing interest in current-induced state transitions in electronic devices, a systematic investigation of Joule heating in electronic devices and the development of comprehensive guidelines for the calculation of temperature changes are imperative and timely. \par

In the field of spintronics, current-driven magnetization switching in antiferromagnets has been studied extensively for the past decade. Antiferromagnets are a class of magnetic materials in which individual magnetic moments order in such a way that there is no net magnetization. Due to their insensitivity to external perturbations and the potential to achieve high storage densities and fast operating speeds, antiferromagnetic materials are considered as promising candidates for next-generation spintronic devices~\cite{jungwirth_antiferromagnetic_2016, baltz_antiferromagnetic_2018, zelezny_spin_2018, siddiqui_metallic_2020}. \par

The use of antiferromagnets for information devices has been challenging due to the difficulty of detecting antiferromagnetic order on a macroscopic scale and the local compensation of spin-related electrical and optical responses by different magnetic sublattices. Recent advances, however, have demonstrated electrical manipulation and detection of magnetic order in antiferromagnets due to their characteristic electronic band structures~\cite{nakatsuji_large_2015, wadley_electrical_2016, tsai_electrical_2020, grzybowski_imaging_2017, matalla-wagner_electrical_2019, bodnar_writing_2018, meinert_electrical_2018, cheng_electrical_2020, han_electrical_2024}. In these studies, Joule heating must be carefully considered as it can play a key role in the switching or create a thermal artifacts~\cite{meinert_electrical_2018, chiang_absence_2019, barry_heat_2019}. \par

Current-driven magnetization switching has been demonstrated in $D0_\mathrm{19}$-type non-collinear antiferromagnets, {\ce{Mn3$X$}} ($X$ = Sn, Ge, etc.). These non-collinear antiferromagnets, characterized by a kagome atomic structure and a chiral spin configuration with three sub-lattices, give rise to a non-vanishing Berry curvature in momentum space, characterized by Weyl nodes. Consequently, such antiferromagnets exhibit significant macroscopic electrical and optical responses dependent on their magnetic structure, such as anomalous Hall effect, anomalous Nernst effect, and magneto-optical Kerr effect at room temperature, despite their negligible net magnetization~\cite{nakatsuji_large_2015, higo_anomalous_2018, ikeda_anomalous_2018,ikhlas_large_2017, higo_large_2018}.  \par

The positions of the Weyl nodes are influenced by the orientation of a cluster magnetic-multipole, {\em i.e.}, an octupole moment, which is aligned with a weak net magnetic moment in the case of {\ce{Mn3Sn}} or {\ce{Mn3Ge}}~\cite{suzuki_cluster_2017,kuroda_evidence_2017}. Therefore, the direction of the octupole moment can be controlled by magnetic fields and electric currents due to the interplay between Zeeman energy and spin torques~\cite{nakatsuji_large_2015, xie_magnetization_2022, tsai_electrical_2020}. \par

Since these non-collinear antiferromagnets, such as {\ce{Mn3Sn}} and {\ce{Mn3Ge}}, have relatively low N{\'e}el temperatures, $\lesssim$~430~K, Joule heating can play a critical role. Although it was initially proposed that the octupole switching is primarily driven by the collective spin rotation induced by the spin-orbit torque and an in-plane magnetic field~\cite{tsai_electrical_2020, higo_perpendicular_2022, yoon_handedness_2023}, other studies have emphasized the pivotal role of heat, and found that the switching occurs around the N{\'e}el temperature~\cite{krishnaswamy_time-dependent_2022, pal_setting_2022}.  This issue is a germane problem in spintronics, where current densities are often close to device breakdown~\cite{lee_thermally_2014, bi_thermally_2014, pham_thermal_2018, taniguchi_magnetization_2022, yuan_anomalous_2023}. It is therefore imperative to have a general method for evaluating the Joule-heating effect. \par

In this article, we present a systematic methodology and an analytical model to investigate the thermal contribution to current-driven octupole switching in the non-collinear antiferromagnet, {\ce{Mn3Sn}}. To this end, we varied the effective thermal resistance independently from the electrical resistance of the device, allowing us to unambiguously identify the role of thermal versus electrical effects. Using {\ce{W/Mn3Sn}} films on {\ce{Si/SiO2}} substrates with different {\ce{SiO2}} layer thicknesses, $h_\mathrm{SiO_2}$, we obtained the threshold current density for octupole switching, $j_\mathrm{th}$, which depends on both the {\ce{SiO2}} thickness and the base temperature, $T_0$. We identified the heating temperature at the threshold current density, $T_\mathrm{th}$, and consistently showed that $T_\mathrm{th}$ remains in all cases above the N{\'e}el temperature, $T_\mathrm{N}$. We also developed an analytical model for calculating the Joule-heating temperature, and showed that it quantitatively describes our experimental results. From numerical simulations, we illustrate the reconfiguration of the magnetic octupoles from the demagnetization state. Our results elucidate the significant role of Joule-heating in current-driven octupole switching in {\ce{Mn3Sn}}, and provide a general method for measuring and calculating Joule-heating temperatures that is applicable to broader research areas using micro-electronic devices. \par

\section{Results and Discussion}
\subsection*{Field-driven octupole switching.}
Thermally oxidized \ce{Si/SiO2} substrates were prepared with different \ce{SiO2} thicknesses, $h_\mathrm{SiO_2}$, ranging from 100 nm to 1000 nm. A \ce{W}(7.1~nm) / \ce{Mn3Sn}(34.4~nm) / \ce{MgO}(2~nm) film was deposited on these substrates at room temperature by dc magnetron sputtering. The samples were annealed at $\SI{500}{\degreeCelsius}$ for 30 minutes. The films are polycrystalline and the atomic ratio of \ce{Mn3Sn} is \ce{Mn}:\ce{Sn} $\approx$ 77:23 [see Supplementary Material]. The excess Mn atoms stabilize the hexagonal structure of \ce{Mn3Sn}~\cite{kren_study_1975}. Note that the thickness of the \ce{SiO2} layer, $h_\mathrm{SiO_2}$, changes the thermal property of the substrates because the thermal conductivity of \ce{SiO2}, $\Lambda_\mathrm{SiO_2} \approx$ \SI{1.3}{\watt\meter^{-1}\kelvin^{-1}}, is about two orders of magnitude smaller than that of \ce{Si}, $\Lambda_\mathrm{Si} \approx $ \SI{140}{\watt\meter^{-1}\kelvin^{-1}}. \par

%
\begin{figure}[t!]
\centering
\includegraphics[width = 7.4cm]{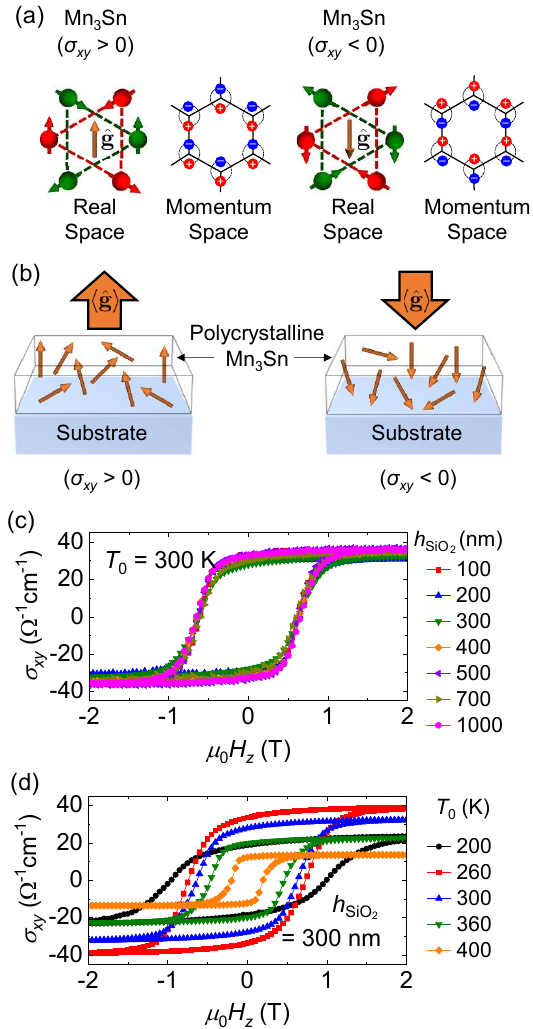}
\caption{\label{Fig:Field_Driven}
    Octupole-moment switching in polycrystalline \ce{Mn3Sn}. (a) Schematics of the magnetic dipole moment configurations in \ce{Mn3Sn} when the octupole moment, $\hat{\mathbf{g}}$, is oriented in the up (left) and down (right) directions. The red and green spheres indicate \ce{Mn} atoms on the first and second kagome layers, respectively. The insets shows schematics of the locations of the Weyl nodes in momentum space near the Fermi level~\cite{kuroda_evidence_2017}. The red and blue circles in the insets represent the positive (+) and negative (-) chirality of the Weyl nodes, respectively. (b) $\hat{\mathbf{g}}$ distribution in a polycrystalline \ce{Mn3Sn} layer when the average of the octupole moment, $\langle \hat{\mathbf{g}} \rangle$, is up and down, in which the anomalous Hall conductivity becomes, $\sigma_{xy} > 0$ and $\sigma_{xy} < 0$, respectively. (c) Hall conductivity, $\sigma_{xy}$, as a function of out-of-plane magnetic field, $H_z$, at 300~K for different thicknesses of \ce{SiO2} layer, $h_\mathrm{SiO_2}$. (d) $\sigma_{xy}$ as a function of $H_z$ measured at different base temperatures, $T_0$, for $h_\mathrm{SiO_2}$ = 300 nm.
}
\end{figure}

We first measured the anomalous Hall conductivity of \ce{W/Mn3Sn}, which depends on the positions of the Weyl nodes in momentum space~\cite{kuroda_evidence_2017}. Since the Weyl points correlate with the spin ordering in \ce{Mn3Sn}, the Hall signal can be manipulated by controlling the octupole-moment direction, $\hat{\mathbf{g}}$ [Fig.~\ref{Fig:Field_Driven}(a)]. The anomalous Hall effect is maximized when $\hat{\mathbf{g}}$ is perpendicular to the film plane, and in the case of polycrystalline \ce{Mn3Sn}, the signal amplitude depends on the average of the out-of-plane component of the octupole moment over the sample, $\langle \hat{\mathbf{g}} \rangle$ [Fig.~\ref{Fig:Field_Driven}(b)]. \par

In Fig.~\ref{Fig:Field_Driven}(c), we plot the Hall conductivity, $\sigma_{xy}$, as a function of the perpendicular magnetic field, $H_z$, for different $h_\mathrm{SiO_2}$ at $T$ = 300 K. As can be seen in Fig.~\ref{Fig:Field_Driven}(c), all samples show similar responses to $H_z$ regardless of $h_\mathrm{SiO_2}$. The residual Hall conductivity is $\sigma_{xy,0} \approx$ \SI{31}{\ohm^{-1} \centi\meter^{-1}}, which is comparable to the previously reported value, $\sim$\SI{20}{\ohm^{-1} \centi\meter^{-1}}, in a single-layer polycrystalline \ce{Mn3Sn} film~\cite{tsai_electrical_2020}. $\sigma_{xy,0}$ of polycrystalline \ce{Mn3Sn} is lower than that of epitaxially grown \ce{W/Mn3Sn}, $\sim$\SI{40}{\ohm^{-1} \centi\meter^{-1}} ~\cite{higo_perpendicular_2022}. The coercive field strength, $\mu_0 H_\mathrm{c}$ = \SI{0.6}{\tesla}, is about three-times larger than that of epitaxially grown \ce{Mn3Sn} films, $\mu_0 H_\mathrm{c} \approx $ \SI{0.2}{\tesla}~\cite{takeuchi_chiral-spin_2021, higo_perpendicular_2022}. \par

We also obtained the hysteresis curves at different base temperatures, $T_0$, ranging from 200~K to 400~K. Note that \ce{Mn3Sn} changes the phase from the non-collinear state to the incommensurate spin-spiral state below 275~K~\cite{duan_magnetic_2015, sung_magnetic_2018}. As shown in Fig.~\ref{Fig:Field_Driven}(d), the shape of the hysteresis loops depends on $T_0$. $\sigma_{xy,0}$ increases with increasing $T_0$ from 200~K to 260~K then decreases with increasing $T_0$. $\sigma_{xy,0}$ becomes zero when $T_0$ reaches the N{\'e}el temperature, $T_\mathrm{N} \approx$ 410~K. The coercivity decreases monotonically with increasing $T_0$ due to thermal fluctuations, which leads to a decrease of the effective magnetic anisotropy. \par

\subsection*{Current-driven octupole switching.}
\begin{figure}[t!]
\centering
\includegraphics[width = 7.4cm]{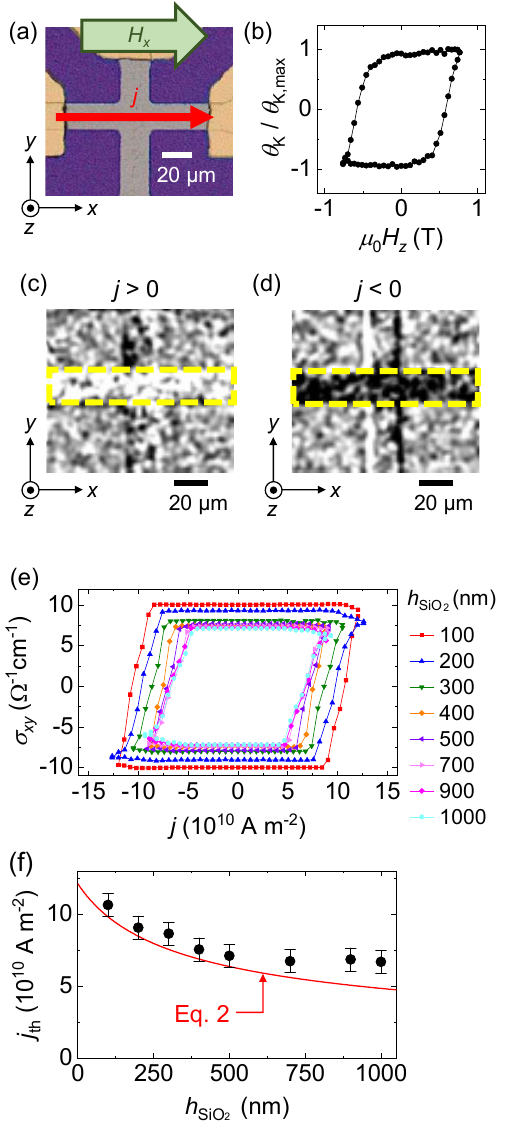}
\caption{\label{Fig:Current_Driven}
    Current-driven octupole switching in \ce{Mn3Sn}. (a) An optical microscopy image of the Hall bar. $j$ and $H_x$ indicate a write current pulse and an in-plane dc magnetic field, respectively. (b) The Kerr rotation angle, $\theta_\mathrm{K}$, as a function of a perpendicular magnetic field, $H_z$. (c)-(d) Magneto-optical Kerr effect microscopy images after applying a positive ($j > 0$) and a negative ($j < 0$) pulse. Yellow rectangles indicate the current flow regions. The brightness reflects the magneto-optical Kerr effect amplitude. To enhance the contrast, the background image before the current application was subtracted and a contrast enhancement technique was used. (e) Hysteresis curves obtained using a current pulse, $j$, for different thicknesses of the \ce{SiO2} layer, $h_\mathrm{SiO_2}$. (f) Threshold current density, $j_\mathrm{th}$ as a function of $h_\mathrm{SiO_2}$. The dots are experimental data and the red line is the calculated $j_\mathrm{N}$ from Eq.~\ref{Eq:jN}.
}
\end{figure}

Subsequently, we measured the current-driven octupole-moment switching. Similar to the case of heavy-metal/ferromagnet structures, the octupole moments can be controlled deterministically by the spin-orbit torque from the adjacent heavy-metal layer~\cite{tsai_electrical_2020}. Note that the contribution of the intergrain spin-transfer torque to the switching is negligible because it is much smaller compared to the spin-orbit torque~\cite{xie_magnetization_2022,xie_efficient_2023}. To generate the spin-orbit torque, we applied a dc electric pulse to \ce{W/Mn3Sn} with a current density $j$ [red arrow in Fig.~\ref{Fig:Current_Driven}(a)], and simultaneously applied a small in-plane dc magnetic field, $H_x$, in the current direction [green arrow in Fig.~\ref{Fig:Current_Driven}(a)]. The pulse duration time and magnetic field strength were 100~ms and 100~mT, respectively. The fall time of the pulse is sufficiently long, $\sim$\SI{1}{\milli\second}, so that we can avoid multi-stable octupole switching~\cite{krishnaswamy_time-dependent_2022}. $\mu_0 H_x =$ \SI{100}{\milli\tesla} is below the coercivity, $\mu_0 H_\mathrm{c}$ = \SI{0.6}{\tesla}, and the value is optimized to maximized the switching efficiency~\cite{pal_setting_2022}. The final effective octupole direction is determined by the directions of the spin-orbit torque and the magnetic field. After injecting the write current pulse, we measured $\sigma_{xy}$ with a sufficiently small read current density, $\sim 1.2 \times 10^9$~\SI{}{\ampere\meter^{-2}}, to detect the octupole state in \ce{Mn3Sn}. \par

%

In addition, we observed the octupole switching using the magneto-optical Kerr effect. First, we measured the Kerr rotation angle, $\theta_\mathrm{K}$, as a function of $H_z$ [Fig.~\ref{Fig:Current_Driven}(b)]. The coercivity in this hysteresis loop is about 0.6 T which corresponds to the coercivity obtained from the anomalous Hall effect in Fig.~\ref{Fig:Field_Driven}(c). The images in Figs.~\ref{Fig:Current_Driven}(c) and \ref{Fig:Current_Driven}(d) show the differential magneto-optical Kerr effect microscopy images after applying a positive ($j > 0$) and a negative ($j < 0$) pulses with positive $H_x$, respectively. The brightness in the current path changes when the pulse direction is reversed, while the outside brightness is almost conserved. These Kerr images clearly show the current-driven octupole switching. \par

We measured the hysteresis loops of the current-driven octupole switching for different $h_\mathrm{SiO_2}$ and obtained the threshold current density, $j_\mathrm{th}$, for the switching [see Figs.~\ref{Fig:Current_Driven}(e) and \ref{Fig:Current_Driven}(f)]. Note that we used the total thickness of the film, $h_\mathrm{f} = 41.5$ nm, including both \ce{W} and \ce{Mn3Sn} layers to calculate the current density, $j$. In contrast to the field-driven switching, $j_\mathrm{th}$ largely decreases with increasing $h_\mathrm{SiO_2}$. $j_\mathrm{th}$ decreases by $\sim$40\% as the thickness of $h_\mathrm{SiO_2}$ increases from 100~nm to 1000~nm. This result shows that the substrate choice plays a crucial role in spin-orbit-torque-driven octupole switching. \par

\subsection*{Temperature excursion due to Joule heating.}

\begin{figure}[t!]
\centering
\includegraphics[width = 7.4cm]{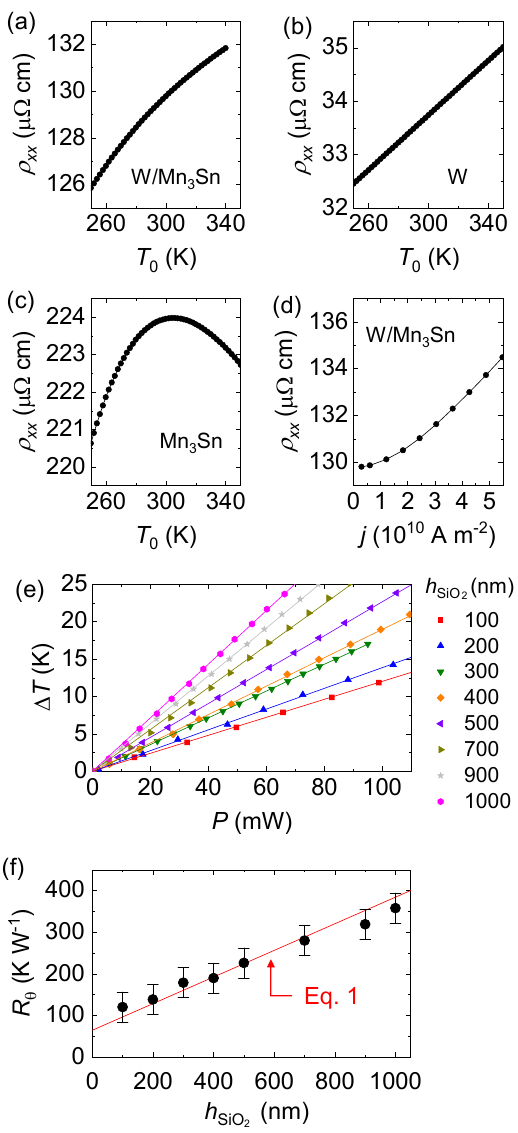}
\caption{\label{Fig:Temp_Inc}
    Effective thermal resistance of \ce{Si/SiO2} substrates. (a)-(c) Longitudinal resistivities, $\rho_{xx}$, of \ce{W/Mn3Sn}, \ce{W}, \ce{Mn3Sn} films as a function of base temperature, $T_0$. (d) $\rho_{xx}$ as a function of current density, $j$, at $T_0$ = 300~K. In (a)-(d), $h_\mathrm{SiO_2}$ = \SI{500}{\nano\meter}. (e) Temperature excursion, $\Delta T$, as a function of input power, $P$, for different $h_\mathrm{SiO_2}$. The symbols and lines indicate the experimental data and the linear fit, respectively. (f) Effective thermal resistance, $R_\theta$, as a function of $h_\mathrm{SiO_2}$. The red line is calculated from Eq.~\ref{Eq:Rth} with $\Lambda_\mathrm{Si}$ = \SI{140}{\watt\meter^{-1}\kelvin^{-1}} and $\Lambda_\mathrm{SiO_2}$ = \SI{1.3}{\watt\meter^{-1}\kelvin^{-1}}.
}
\end{figure}

We investigated the temperature excursion due to Joule heating for different $h_\mathrm{SiO_2}$ to identify the switching temperature. At steady state, where the generated heat balances the heat loss to the substrate, the temperature increase, $\Delta T$, in the \ce{W/Mn3Sn} microstrip is proportional to the input power, $P$, {\em i.e.}, $\Delta T = R_\mathrm{\theta} P$, where $R_\mathrm{\theta}$ is the effective thermal resistance of a substrate. It is important to note that the temperature approaches a saturated state at 100~ms because the heat diffusion distance in silicon, $\sim$2~mm, is much larger than any length scale in our device, $\lesssim$\SI{100}{\micro\meter} [see Supplementary Material]. \par

To measure $R_\mathrm{\theta}$ for different $h_\mathrm{SiO_2}$, we first determined the longitudinal resistivity, $\rho_{xx}$, of \ce{W/Mn3Sn} as a function of the base temperature, $T_0$. A small electrical measurement current density, $\sim 1.2 \times 10^9$~\SI{}{\ampere\meter^{-2}}, was applied to minimize the temperature increase due to Joule heating, $<$1~K.  Figure~\ref{Fig:Temp_Inc}(a) shows a representative $\rho_{xx}$-$T$ relationship for $h_\mathrm{SiO_2}$ = 500~nm. $\rho_{xx}$ of the \ce{W/Mn3Sn} film increases monotonically, but the slope gradually decreases because $\rho_{xx}$ of \ce{W} and \ce{Mn3Sn} have different temperature dependencies. For \ce{W}, $\rho_{xx}$ is mostly proportional to $T_0$, as is typical for conventional metals, whereas \ce{Mn3Sn} exhibits a non-linear response. In particular, $\rho_{xx}$ of \ce{Mn3Sn} increases with increasing temperature below 300~K and then decreases moderately [Figs.~\ref{Fig:Temp_Inc}(b) and \ref{Fig:Temp_Inc}(c)]. \par

We also measured $\rho_{xx}$ as a function of $j$ at room temperature [Fig.~\ref{Fig:Temp_Inc}(d)]. For this measurement, we applied a 100~ms current pulse, similar to those used in current-driven switching, and observed the maximum resistivity before the end of the pulse. As shown in Fig.~\ref{Fig:Temp_Inc}(d), $\rho_{xx}$ increases with increasing $j$ due to Joule heating. Note that $\rho_{xx}$ largely follows a quadratic function of $j$, but there are deviations because $\rho_{xx}$ is not strictly proportional to $T_0$. \par

Using the relationships of $\rho_{xx}$ - $T_0$ and $\rho_{xx}$ - $j$ in Figs.~\ref{Fig:Temp_Inc}(a) and \ref{Fig:Temp_Inc}(d), we determined the temperature increase in \ce{W/Mn3Sn} due to Joule heating, $\Delta T$. The results are plotted as a function of $P$ for different $h_\mathrm{SiO_2}$ in Fig.~\ref{Fig:Temp_Inc}(e). The plots show that $\Delta T$ is proportional to $P$ and the heating temperature depends on $h_\mathrm{SiO_2}$. We obtained the effective thermal resistance, $R_\mathrm{\theta}$, from the slope of the linear fits and plotted $R_\mathrm{\theta}$ as a function of $h_\mathrm{SiO_2}$ in Fig.~\ref{Fig:Temp_Inc}(f) (dot symbols). We observed that $R_\mathrm{\theta}$ increases monotonically with increasing $h_\mathrm{SiO_2}$. \par

%
\begin{figure}[t!]
\centering
\includegraphics[width = 7cm]{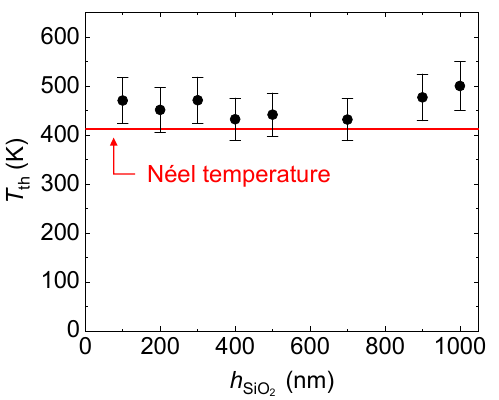}
\caption{\label{Fig:Threshold_Temp}
    Temperature, $T_\mathrm{th}$, at threshold current density as a function of $h_\mathrm{SiO_2}$. The red line shows the N{\'e}el temperature, $T_\mathrm{N} \approx$ 410~K.
}
\end{figure}

From $R_\mathrm{\theta}$ and $P_\mathrm{th}$, we calculated the threshold temperature, $T_\mathrm{th} = T_0 + R_\mathrm{\theta} P_\mathrm{th}$, for various $h_\mathrm{SiO_2}$ [Fig.~\ref{Fig:Threshold_Temp}]. Here, $P_\mathrm{th}$ represents the threshold input power obtained from $j_\mathrm{th}$ and $\rho_{xx} (j)$. $T_\mathrm{th}$ is mostly independent of $h_\mathrm{SiO_2}$, and the average $T_\mathrm{th}$ is about \SI{460}{\kelvin}, which is higher than the N{\'e}el temperature, $T_\mathrm{N} \approx$ \SI{410}{\kelvin}. This result shows that temperature plays an important role in the octupole switching of \ce{Mn3Sn}, and the switching requires temperatures above the N{\'e}el temperature. Note that the obtained {$T_\mathrm{th}$} in Fig.~{\ref{Fig:Threshold_Temp}} is the threshold temperature in the electrode, and the actual temperature at the center of the Hall bar cross is lower than {$T_\mathrm{th}$} due to the heat and current distribution at the cross structure.\par

%
\begin{figure}[t!]
\centering
\includegraphics[width = 7.4cm]{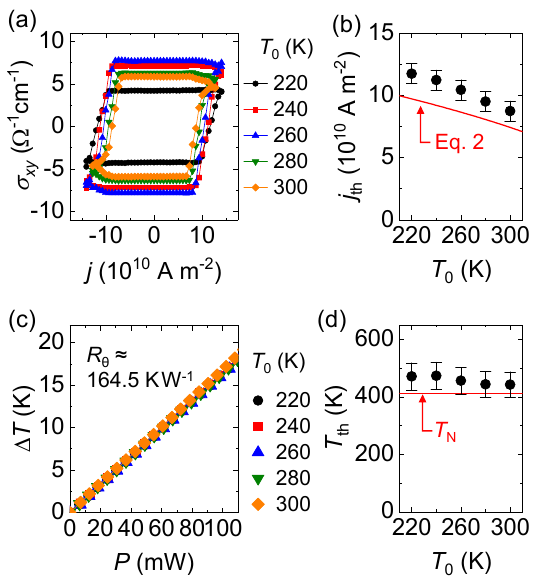}
\caption{\label{Fig:Low_Temp}
    Current-driven octupole switching below room temperature. (a) Hall conductivity, $\sigma_{xy}$, as a function of write current density, $j$, under different base temperatures, $T_0$. The \ce{SiO2} thickness is 300~nm. (b) Threshold current density, $j_\mathrm{th}$, as a function of $T_0$. The dots are experimental data and the solid line is the calculated $j_\mathrm{N}$ from Eq.~\ref{Eq:jN} (c). Temperature excursions due to Joule heating, $\Delta T$, as a function of input power, $P$, under different $T_0$. The symbols and lines indicate the experimental data and the linear fit, respectively. The slope corresponds to the effective thermal resistance, $R_\mathrm{\theta}$ which is about \SI{164.5}{\kelvin\watt^{-1}} for all $T_0$ (d) Threshold temperature, $T_\mathrm{th}$, as a function of $T_0$. The red line shows the N{\'e}el temperature, $T_\mathrm{N} \approx$ 410~K.
}
\end{figure}
We also measured $T_\mathrm{th}$ below room temperature. As shown in Fig.~\ref{Fig:Field_Driven}(d), $\sigma_{xy,0}$ peaks around $T_0 \approx$260~K, then decreases with decreasing temperature due to a phase transition~\cite{duan_magnetic_2015}. Therefore, here we measured $T_\mathrm{th}$ between 220~K and 300~K. As shown in Fig.~\ref{Fig:Low_Temp}(a), the hysteresis behavior varies with $T_0$. In particular, $j_\mathrm{th}$ decreases monotonically with increasing $T_0$ [Fig.~\ref{Fig:Low_Temp}(b)]. We measured $\Delta T$ as a function of $P$ at different $T_0$, and obtained $R_\mathrm{\theta} \approx$~\SI{164.5}{\kelvin \watt^{-1}} [Fig.~\ref{Fig:Low_Temp}(c)], which remains constant in this temperature range. By combining $j_\mathrm{th}$ and $R_\mathrm{\theta}$, we calculated $T_\mathrm{th}$ [Fig.~\ref{Fig:Low_Temp}(d)]. $T_\mathrm{th}$ is independent of $T_0$, averaging $T_\mathrm{th}$ is about 460 K, consistent with the average $T_\mathrm{th}$ in Fig.~\ref{Fig:Threshold_Temp}. These consistent results confirm the reliability of our temperature measurement method. \par

\subsection*{Analytical model to calculate the temperature excursion due to Joule heating.}

In Figs.~\ref{Fig:Threshold_Temp} and \ref{Fig:Low_Temp}(d), we show that the switching temperature, $T_\mathrm{th}$, remains always above the N{\'e}el temperature over all substrate choices and base temperatures. This result implies that temperature plays an important role in current-driven octupole switching in \ce{Mn3Sn} films. Assuming that switching occurs when the temperature reaches a certain threshold, $\sim T_{\mathrm{N}}$, the dependence of $j_\mathrm{th}$ on $h_{\mathrm{SiO_2}}$ and $T_0$ can be explained by the current density required to reach the N{\'e}el temperature, $j_\mathrm{N}$. \par

We calculated the current density required to reach $T_{\mathrm{N}}$. Let us consider a conductive strip placed on a sufficiently wide \ce{Si/SiO2} substrate. The film thickness, $h_\mathrm{f}$, is significantly thinner than the width of the strip, $w$, i.e., $h_\mathrm{f} \ll w$. As soon as an electric current is applied, the temperature increases due to Joule heating and stabilizes after a while within 100~ms, because 100~ms is about $10^3$ times longer than the characteristic thermal time scale in the device, $\sim$0.1~ms. [Supplementary Material]. In this study, we define this condition as steady state, where the temperature rise can be calculated by $\Delta T = R_\mathrm{\theta} P$ which is independent of time. The effective thermal resistance of a \ce{Si/SiO2} substrate,  $R_\mathrm{\theta}$, can be obtained from  the thermal resistance of the \ce{SiO2} layer, $R_{\theta, \mathrm{SiO_2}}$, and \ce{Si} substrate, $R_\mathrm{\theta, \mathrm{Si}}$. \par

First, we calculate the effective thermal resistance of the thin \ce{SiO2} layer, $R_{\theta, \mathrm{SiO_2}}$. When the \ce{SiO2} layer is very thin, $h_\mathrm{SiO_2} \ll w$, the generated heat mostly flows perpendicular to the film plane, and the lateral heat flow can be neglected. In this case, $R_{\theta, \mathrm{SiO_2}}$ can be calculated by $R_{\theta, \mathrm{SiO_2}} = h_\mathrm{SiO_2} /(w l \Lambda_\mathrm{SiO_2})$ at steady state, where $\Lambda_\mathrm{SiO_2}$ is the thermal conductivity of \ce{SiO2} and $l$ is the length of the conductor. Note that the effect of contact pads on the electrode is excluded here because Joule heating in these regions is drastically reduced due to the low current density. \par

Next, we consider the effective thermal resistance of the bulk \ce{Si} substrate, $R_\mathrm{\theta, \mathrm{Si}}$. In a two-dimensional model with a long electrode on a large substrate, $R_\mathrm{\theta, \mathrm{Si}}$ is given by $R_\mathrm{\theta, \mathrm{Si}} = \ln(\eta l_\mathrm{d}/w)/(\pi \Lambda_\mathrm{Si} l)$, where $l_\mathrm{d} = \sqrt{Dt_\mathrm{p}}$ is the thermal diffusion length, $\eta$ is a constant, $D = \Lambda/(\rho C)$ is the thermal diffusivity, $\rho$ is the density, and $C$ is the specific heat~\cite{cahill_thermal_nodate}. However, in our system, $R_\mathrm{\theta, \mathrm{Si}}$ is almost saturated within $\sim$~\SI{1}{\milli\second}, which is much shorter than the pulse length, \SI{100}{\milli\second} [Supplementary Material]. Therefore, $R_\mathrm{\theta, \mathrm{Si}}$ is considered as a time-independent constant in this study. \par

Since the two-dimensional model is not suitable for our system, we numerically calculated the effective thermal resistance of a \ce{Si} substrate at steady state, $R_\mathrm{\theta, \mathrm{Si}}$, considering different values of $w$, $D$, and $l$ [Supplementary Material]. From the simulations, we extrinsically obtained an analytical form, $R_\mathrm{\theta, \mathrm{Si}} = \ln{(\eta' l/w)}/(\pi \Lambda_\mathrm{Si} l)$, where $\eta'$ is about 5 when $l \ll h_\mathrm{Si}$, where $h_\mathrm{Si}$ is the thickness of the substrate. In our experiments, $l$ = \SI{120}{\micro\meter}, $w$ = \SI{20}{\micro\meter}, and $h_\mathrm{Si}$ = \SI{500}{\micro\meter}. \par

By combining $R_\mathrm{\theta, SiO_2}$ and $R_\mathrm{\theta, Si}$, we can calculate the effective thermal resistance of \ce{Si/SiO2}, $R_\mathrm{\theta}$,
\begin{equation} \label{Eq:Rth}
\begin{split}
	R_\mathrm{\theta} = \frac{h_\mathrm{SiO_2}}{\Lambda_\mathrm{SiO_2} l w} + \frac{\ln(\eta' l/w)}{\pi \Lambda_\mathrm{Si} l}.
\end{split}
\end{equation}
We calculate Eq.~\ref{Eq:Rth} with $\Lambda_\mathrm{Si}$ = \SI{140}{\watt\meter^{-1}\kelvin^{-1}}, $\Lambda_\mathrm{SiO_2}$ = \SI{1.3}{\watt\meter^{-1}\kelvin^{-1}}, and plot the results in Fig.~\ref{Fig:Temp_Inc}(f) (red line), which shows quantitatively good agreement with our experimental data. From $\Delta T~=~R_\theta P$ and Eq.~\ref{Eq:Rth}, the current density required to reach the N{\'e}el temperature, $j_\mathrm{N}$, can be calculated as follows
\begin{equation} \label{Eq:jN}
\begin{split}
	j_\mathrm{N}^2 &= \frac{T_\mathrm{N}-T_0}{ \rho_{xx} h_\mathrm{f} w l R_{\theta}}.
\end{split}
\end{equation}
$j_\mathrm{N}$ in Eq.~\ref{Eq:jN} is plotted in Figs.~\ref{Fig:Current_Driven}(f) and ~\ref{Fig:Low_Temp}(b). The calculated $j_\mathrm{N}$ is about 20\% lower than $j_\mathrm{th}$, but it describes well the dependence of $j_\mathrm{th}$ on both $h_\mathrm{SiO_2}$ and $T_0$. In all cases, there is a quantitatively and qualitatively reasonable agreement without any free fitting parameters. Note that the discrepancy between the calculated $j_\mathrm{N}$ and the measured $j_\mathrm{th}$ can be attributed to the differences between the measured threshold temperature, $T_\mathrm{th}$, and the N{\'e}el temperature, $T_\mathrm{N}$, due to the reduction of the current density at the center of the Hall bar cross, as shown in Figs.~\ref{Fig:Threshold_Temp} and \ref{Fig:Low_Temp}(d). \par

To confirm the universality of our model, we calculated the threshold current density using {Eq.~\ref{Eq:jN}} in the case of 100\% octupole configuration switching in epitaxial {\ce{W/Mn3Sn}} heterostructures~{\cite{higo_perpendicular_2022}}. In this study, the threshold current density is approximately {$7.5 \times 10^{10}$~\SI{}{\ampere\meter^{-2}}}. Using appropriate parameters, {$\Lambda_\mathrm{MgO}$} = {\SI{40}{\watt\meter^{-1}\kelvin^{-1}}}, {$T_\mathrm{th}$} = 460~K, {$\rho_{xx}$} = {\SI{81}{\micro\ohm\centi\meter^{-1}}}, {$h_\mathrm{f}$} = 37~nm, {$w$} = {\SI{32}{\micro\meter}}, and {$l$} = {\SI{200}{\micro\meter}}, we obtain {$j_\mathrm{N} \approx 7.9 \times 10^{10}$}~{\SI{}{\ampere\meter^{-2}}}, which is in good agreement with their experimental results. This calculation result implies that temperature also plays a crucial role in octupole switching in epitaxially grown {\ce{W/Mn3Sn}} films. \par

\subsection*{Numerical calculations of the octupole reconfiguration from the demagnetization state.}

\begin{figure}[t!]
\centering
\includegraphics[width = 7.4 cm]{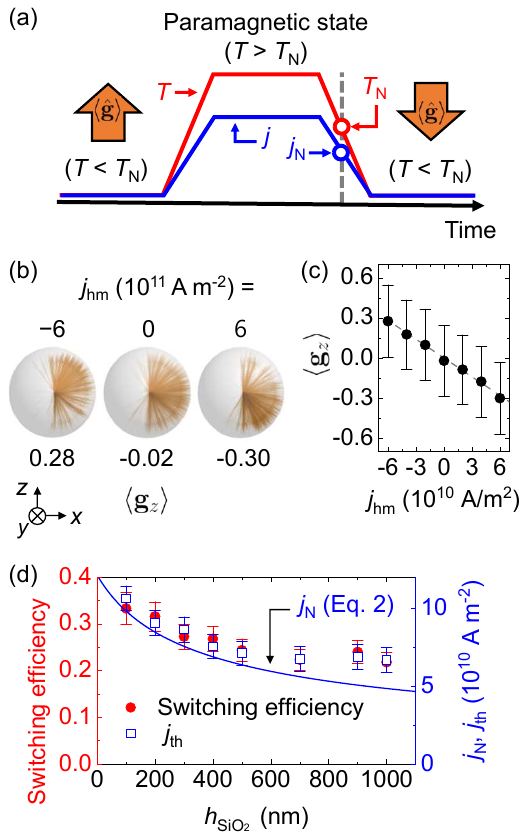}
\caption{\label{Fig:Numeric_Calc}
    Octupole reconfiguration from demagnetization state in polycrystalline \ce{Mn3Sn}. (a) Schematics of the thermal octupole switching process. The blue and red lines represent the time evolution of a current pulse, $j$, and temperature, $T$, respectively. The dashed line indicates the time when the temperature decreases below the N{\'e}el temperature, where the octupole formation starts. The orange arrows show the averaged octupole polarization, $\langle \hat{\mathbf{g}} \rangle$. (b) Calculated final-octupole orientations from a thousand random models under different current density, $j$. Both the initial magnetic moments and the crystal orientations are randomly distributed [Supplementary Material]. The positive directions of a magnetic field and the spin torque are parallel to the $+x$ and $-y$-directions, respectively. (c) The averaged $z$-component of the octupole moment, $\langle \mathbf{g}_z \rangle$, as a function of $j$. The dashed line is a linear fit. (d) The switching efficiency (red dots and left $y$-axis), the currenct density at the N{\'e}el temperature, $j_\mathrm{N}$, (blue line and right $y$-axis), and threshold current density, $j_\mathrm{th}$ (blue squares and right $y$-axis) as a function of the thickness of \ce{SiO2} layers, $h_\mathrm{SiO_2}$.
}
\end{figure}

So far, we have demonstrated the importance of temperature in current-driven octupole switching in \ce{Mn3Sn} through systematic experiments and an
analytical model to evaluate the Joule heating effects~\cite{krishnaswamy_time-dependent_2022,pal_setting_2022}. When an electric pulse with $j > j_\mathrm{N}$ is applied, the temperature rises above $T_{\mathrm{N}}$, leading to the disappearance of the octupole, while the individual magnetic moments exhibit random dynamics in the paramagnetic state [Fig.~\ref{Fig:Numeric_Calc}(a)]. As the current decreases after the pulse, the temperature falls below $T_{\mathrm{N}}$, and the magnetic moments begin to rearrange from a random orientation. Temperature alone does not provide a preference for the octupole orientation. Instead, the averaged octupole orientation is influenced by the residual spin torques below $T_\mathrm{N}$. Octupole re-configuration is initiated near the interface between \ce{W} and \ce{Mn3Sn} due to the short spin-diffusion length in \ce{Mn3Sn}, $\sim$~\SI{1}{\nano\meter}, which seeds the spin texture for the entire layer~\cite{pal_setting_2022}. \par

%

To investigate the control of the octupole state from the demagnetization state, we performed numerical calculations [Methods and Supplementary Material]. We computed the spin dynamics of a thousand unit cells with random crystal orientations and random initial spin configurations to consider the polycrystalline structure and the demagnetization state just below $T_\mathrm{N}$, respectively. For simplicity, we have not considered the self-induced spin-transfer torque in the polycrystalline structure because its effect is much smaller than that of the spin-orbit torque.~\cite{xie_magnetization_2022,xie_efficient_2023}. \par

Once the simulation starts, the octupole is rapidly formed from a demagnetized state in $\sim$\SI{0.1}{\nano\second}, followed by the rotational dynamics of the octupole moment [Supplementary Material]. The octupole motion mostly stops at the energy minimum state in a few ns. The final octupole orientation is determined by the spin-orbit torque, the magnetic field, and the crystal orientation. In these calculations, we assumed that the octupole dynamics occur under nearly constant current and temperature just below $T_\mathrm{N}$, because the fall time of the current, $\sim$\SI{1}{\milli\second}, is much longer than the time for the octupole dynamics, $\lesssim$~10 ns. \par

The distributions of the final octupole orientations are shown in Fig.~\ref{Fig:Numeric_Calc}(b) for different current densities, $j_\mathrm{hm}$ = $-6 \times 10^{11}$, 0, and $6 \times 10^{11}$ \SI{}{\ampere\meter^{-2}} where $j_\mathrm{hm}$ represents the current density in the heavy-metal layer. These orientations are obtained in the ground state after both current and field are turned off. When no current is applied, {\em i.e.,} $j_\mathrm{hm}$ = 0, the octupole directions are mainly oriented in the $+x$ direction due to the magnetic field, and are uniformly distributed in the $z$-direction, resulting in an almost zero average out-of-plane octupole moment, $\langle \mathbf{g}_z \rangle$. In the presence of an electric current, however, the octupoles show a preference for certain orientations depending on the current direction. Positive and negative currents induce more negative and more positive $\langle \mathbf{g}_z \rangle$, respectively. Figure~\ref{Fig:Numeric_Calc}(b) shows that octupole orientations can be determined stochastically by the current direction in the polycrystalline \ce{Mn3Sn}. \par

Figure~\ref{Fig:Numeric_Calc}(c) shows that $\langle \mathbf{g}_z \rangle$ is proportional to $j$, and this implies that the switching efficiency is proportional to $j_\mathrm{th}$ and $j_\mathrm{N}$, because the octupole formation occurs just below $T_\mathrm{N}$. The threshold current density, $j_\mathrm{N}$, can be obtained from Eq.~\ref{Eq:jN} by assuming a quasi-static state. To confirm the relation between $j_\mathrm{th}$ and switching efficiency, we plotted the switching efficiency, $j_\mathrm{N}$, and $j_\mathrm{th}$ as a function of $h_\mathrm{SiO_2}$ in Fig.~\ref{Fig:Numeric_Calc}(d). The switching efficiency is obtained from $\sigma_{xy,0}^\mathrm{sot} / \sigma_{xy,0}^\mathrm{field}$, where  $\sigma_{xy,0}^\mathrm{field}$ and $\sigma_{xy,0}^\mathrm{sot}$ are the residual anomalous Hall conductivity obtained from field- and current-driven octupole switching, respectively [Figs.~\ref{Fig:Field_Driven}(c) and \ref{Fig:Current_Driven}(e)]. These results confirm that the switching efficiency is proportional to $j_\mathrm{N}$ and $j_\mathrm{th}$, which is consistent with the numerical simulation results in Fig.~\ref{Fig:Numeric_Calc}(c). \par

\section{Summary and outlook}
In this study, we studied a systematic methodology and an analytical model to evaluate the temperature excursion due to Joule heating in spintronic devices based on current-driven octupole switching in the non-collinear antiferromagnet, {\ce{Mn3Sn}}. The results show that the threshold current density for switching depends on both the substrate choice and the base temperature, while the switching temperature remains essentially constant above the N{\'e}el temperature. The calculated current density for reaching the N{\'e}el temperature from an analytical model quantitatively describes the dependence of the switching current density. From numerical calculations, we showed the octupole reconfiguration during cooling in polycrystalline {\ce{Mn3Sn}} from the demagnetization state. This work elucidates the role of temperature in octupole switching in {\ce{Mn3Sn}} and offers fundamental insights into the demagnetization-mediated octupole reconfiguration. In addition, it provides a general guideline for characterizing the Joule-heating temperature in micro-electronic devices, thus allowing a careful assessment of thermal effects, which is often a concern in spintronic devices. \par

It is important to note that, in this study, we focused only on spin-orbit-torque-assisted octupole switching with an adjacent heavy-metal layer. However, it has recently been reported that switching can also be achieved by other spin torques, such as the self-induced spin-transfer torque and the orbital Hall effect, where the role of temperature in octupole switching may be different~{\cite{xie_magnetization_2022,xie_efficient_2023}}. Therefore, further studies with different spin-torque sources are needed. \par


\section{Methods}

\subsection*{Sample preparation and characterization}

Thin films of W~(7.1~nm)/\ce{Mn3Sn}~(34.4~nm)/MgO~(2~nm) were deposited on \ce{Si/SiO2} substrates by dc magnetron sputtering. \ce{SiO2} layers with eight different thicknesses ($h_\mathrm{SiO_2}$ = 100, 200, 300, 400, 500, 700, 900, and 1000~nm) were used in this study . \ce{W}, \ce{Mn3Sn}, and \ce{MgO} films were deposited at room temperature with a chamber base pressure of $\sim 6 \times 10^{-9}$~Torr. The \ce{Mn3Sn} films were co-sputtered from \ce{Mn} and \ce{Sn} targets. The deposition rates of \ce{W} and \ce{Mn3Sn} were approximately 0.09~nm/s and 0.05~nm/s, respectively. After deposition, the films were annealed in situ at \SI{500}{\celsius} for 30 minutes. \par

Rutherford backscattering was used to determine the composition of the \ce{Mn3Sn} film. The stoichiometry of \ce{Mn3Sn} was determined to be close to \ce{Mn}:\ce{Sn} = 77:23 [Supplementary Material]. \par

The film thicknesses were measured by X-ray reflection with a Cu-K$\alpha_1$ X-ray at a wavelength of \SI{0.154}{\nano\meter}. The surface roughness of \ce{W} and \ce{Mn3Sn} is about $\sim$~\SI{0.5}{\nano\meter} and $\sim$~\SI{1}{\nano\meter}, respectively [Supplementary Material]. \par

The magnetization as a function of perpendicular field was measured using a commercial superconducting quantum interference device (SQUID) magnetometer (MPMS3, Quantum Design). The saturation magnetization of \ce{Mn3Sn} is approximately 0.03~$\mu_\mathrm{B}$/f.u. [Supplementary Material]. \par

\subsection*{Transport measurements}

For transport measurements, the films were patterned into Hall bar geometries with $\sim$~\SI{20}{\micro\meter} width using conventional photo-lithography techniques (375-nm maskless laser writer; MLA 150, Heidelberg). The width of the electrodes is about \SI{20}{\micro\meter}, and the distance between two Hall bars is about \SI{120}{\micro\meter}. The length of the electrode for the current-driven switching is $\sim$~\SI{100}{\micro\meter}. For the electrical characterization of the devices, Ti~(5~nm)/Au~(90~nm) contacts were used. \par

In the current-driven switching experiments, we applied a write current pulse with a duration of 100~ms. Subsequently, we applied a small dc current, $\sim1.2 \times 10^9$~\SI{}{\ampere\meter^{-2}}, and measured the Hall voltage. A dc in-plane magnetic field of 0.1~T was applied consistently in the current direction, $+x$-direction, during both the write and read processes. \par

To measure the longitudinal resistivity of the \ce{Mn3Sn} devices, a four-probe method was used. We applied a 100~ms-long current pulse of $\sim1.2 \times 10^9$~\SI{}{\ampere\meter^{-2}} and measured the maximum voltage during the pulse application [Supplementary Material]. \par

The current density, resistivity, and conductivity in the main text are calculated based on the total thickness of W~(7.1~nm)/\ce{Mn3Sn}~(34.4~nm), $h_{\mathrm{f}}$~=~\SI{41.5}{\nano\meter}. The low and high temperature measurements were performed using a commercial physical property measurement system (PPMS, Quantum Design). \par

\subsection*{Magneto-optical Kerr effect measurements}

The Kerr rotation angle was measured using the polar magneto-optical Kerr effect (MOKE) system with a 636 nm laser under an out-of-plane external magnetic field~\cite{polisetty_optimization_2008}. Both incident and reflected angles were nearly perpendicular to the film plane. The incident beam was linearly polarized and the reflected beam was periodically modulated between left and right circularly polarized light by the photoelastic modulator (PEM-100, Hinds Instruments). The modulation is performed at a frequency of 50 kHz and a retardation amplitude of 2.405 radians. The analyzer is set at about \SI{45}{\degree}. The beam transmits through an analyzer and is finally detected by a photosensitive diode which provides the input signal to the lock-in amplifier (Stanford Research Systems, SR830) and a dc voltmeter. \par

Based on the Kerr rotation angle measurement, MOKE imaging was performed using a polar MOKE microscope to observe the current-driven octupole switching. MOKE provides direct visualization of the spatial distribution of the out-of-plane component of the magnetic octupole polarization. To increase the image contrast, we obtained differential images by subtracting the initial image, and a contrast enhancement technique is used by saturating 3\% of the pixels of the images. All MOKE microscopy measurements were performed at room temperature.

\subsection*{Numerical calculations for spin dynamics}

We studied the dynamics of the sub-lattice moments based on the Landau-Lifshitz equation with the spin-orbit torque,
\begin{equation} \label{Eq:LLeq}
\begin{split}
    \frac{d \mathbf{m}_i}{d t} = &-\frac{|\gamma|}{1+\lambda^2} \left[ \mathbf{m}_i \times \mathbf{B}^i_\mathrm{eff} + \lambda \mathbf{m}_i \times \left( \mathbf{m}_i \times \mathbf{B}^i_\mathrm{eff} \right) \right] \\
    &+ \frac{\theta_\mathrm{sh} \hbar |\gamma| j_\mathrm{hm}}{2 e M_\mathrm{s} h} \mathbf{m}_i \times \left( \mathbf{m}_i \times \mathbf{\sigma} \right)
\end{split}
\end{equation}
where $\mathbf{m}_{i}$ is a unit magnetic moment of site $i$, $\gamma$ is the gyromagnetic ratio, $\lambda$ is a damping constant, $\theta_\mathrm{sh}$ is a spin-Hall angle, $e$ is the charge of an electron, $M_\mathbf{s}$ is the saturation magnetic moment of the \ce{Mn} atom, $h$ is a film thickness, $\mathbf{\sigma}$ is a spin polarization vector, and $j_\mathrm{hm}$ is a current density in the heavy-metal layer. $\mathbf{B}^i_\mathrm{eff}$ is the effective magnetic field at each magnetic moment. $\mathbf{B}^i_\mathrm{eff}$ can be calculated as
\begin{equation} \label{Eq:Beff}
\begin{split}
    \mathbf{B}^i_\mathrm{eff} = -\frac{1}{\mu_0 M_\mathrm{s}} \frac{\partial \cal{H}}{\partial \mathbf{m}_i},
\end{split}
\end{equation}
where $\cal{H}$ is the total magnetic energy in a unit cell,
\begin{equation} \label{Eq:Hamiltonian}
\begin{split}
    H = &-J \sum_{\langle ia, jb \rangle} \mathbf{m}_{ia} \cdot \mathbf{m}_{jb} \\
    &- D \sum_{\langle i, j \rangle} \hat{\mathbf{z}} \cdot (\mathbf{m}_{i1} \times \mathbf{m}_{j2}+\mathbf{m}_{i2} \times \mathbf{m}_{j3}+\mathbf{m}_{i3} \times \mathbf{m}_{j1}) \\
    &- K \sum_{ia} (\mathbf{k}_a \cdot \mathbf{m}_{ia})^2-\mu_0 M_\mathrm{s} \sum_{ia} \mathbf{m}_{ia} \cdot \mathbf{H},
\end{split}
\end{equation}
where the suffixes $i$ and $j$ denote a unit cell, and $a$ and $b$ denote the sub-lattices. Our model includes six Mn atoms in a unit cell, taking into account inter-layer couplings [Fig.~\ref{Fig:Field_Driven}(a)]. For this calculation, we used the periodic boundary condition in the $x$-, $y$- and $z$-directions. $J$, $D$, and $K$ are the nearest neighbor exchange interaction, the Dzyaloshinskii-Moriya interaction, and the in-plane magnetic anisotropy, respectively. $\mathbf{H}$ is an external magnetic field. $\mathbf{k}_a = \left( 0, \cos \varphi_a, \sin \varphi_a \right)$ with $(\varphi_1, \varphi_2, \varphi_3) = (\varphi_4, \varphi_5, \varphi_6) = (\pi, 9 \pi, 5 \pi)/6$, where $\varphi_a$ is the angle of the sublattice moments with respect to the $x$ axis in the kagome plane, {\em i.e.,} $x$-$y$ plane. In the numerical calculations, we used $J = -2.8$ meV, $D = -0.22$ meV, and $K$ = 0.187 meV, $\theta_\mathrm{sh} = 0.1$, $M_\mathrm{s}$ = 3 $\mu_0$, and $h$ = \SI{40}{\nano\meter}. We did not considered the strain effect which causes the uniaxial anisotropy.~\cite{park_magnetic_2018, higo_perpendicular_2022}.

\section{Acknowledgements}

This research is supported by the NSF through the University of Illinois Urbana-Champaign Materials Research Science and Engineering Center Grant No. DMR-1720633 and is carried out in part in the Materials Research Laboratory Central Research Facilities, University of Illinois.

%
\bibliographystyle{apsrev4-2}
\bibliography{references}{}

\end{document}


\title{\Large Supplementary Material for\\ ``Thermal contribution to current-driven antiferromagnetic-order switching''\large }

\author{Myoung-Woo Yoo}
\affiliation{\MRL}
\affiliation{\MSE}
\author{Virginia O. Lorenz}
\affiliation{\MRL}
\affiliation{\PHY}
\author{Axel Hoffmann}
\affiliation{\MRL}
\affiliation{\MSE}
\affiliation{\PHY}
\author{David G. Cahill}
\affiliation{\MRL}
\affiliation{\MSE}
\affiliation{\PHY}

%
\maketitle

\section{Temperature increase due to Joule heating}
We performed numerical simulations to determine the effective thermal resistance of a bulk Si substrate, $R_\mathrm{\theta, Si}$, using a commercial simulation software, Ansys Electronics Desktop Student (https://www.ansys.com). \par

In our simulations, we modeled a cuboid-shaped conductor with dimensions of length, $l$, width, $w$, and an ultra-thin height of \SI{0.1}{\micro\meter} positioned on a substrate. The lateral size and thickness of the substrate are $5 \times 5$~\SI{}{\milli\meter^2} and $h_\mathrm{Si}$~=~\SI{500}{\micro\meter}, respectively. We did not include contact pads outside the electrode in the simulation model. The side faces of the substrate were set to ambient temperature, so that the heat generated by the conductor would dissipate primarily through the Si substrate, with minimal transfer to the air via the top and bottom surfaces of the substrate. \par

\begin{figure*}[b!]
\centering
\includegraphics[width = 16cm]{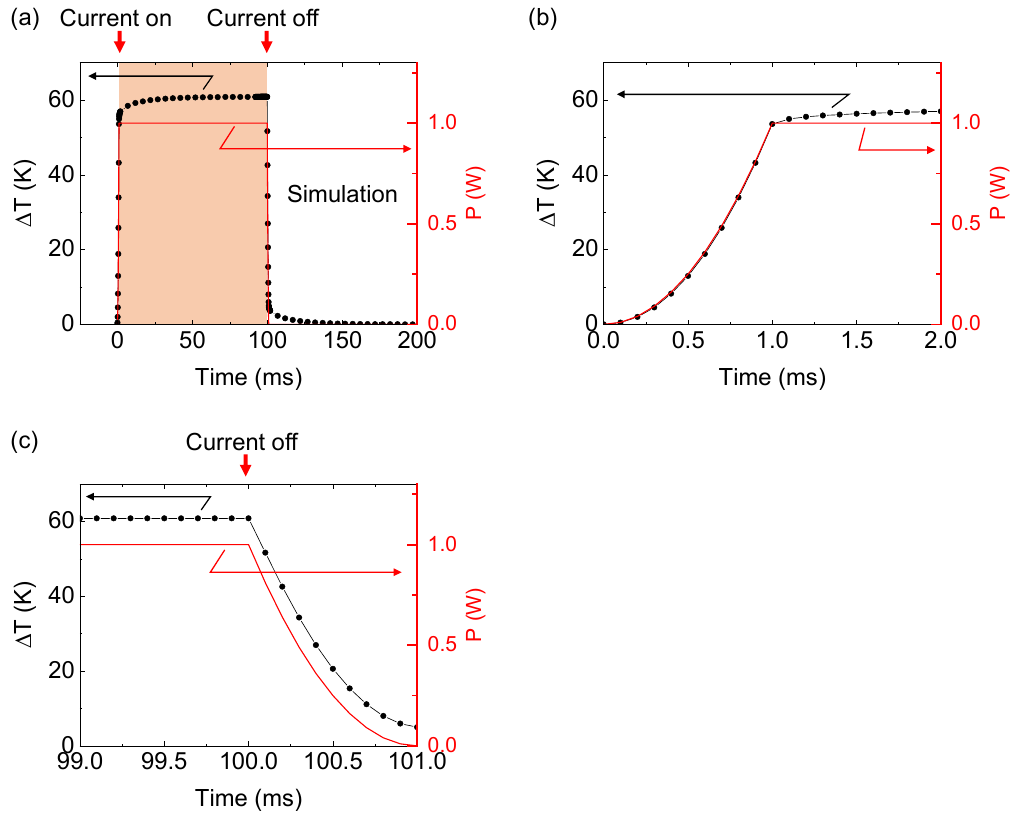}
\caption{\label{Fig:TimeDependent}
    Temperature response to a current pulse. (a) Time evolution of the temperature change, $\Delta T$, during the pulse application (highlighted in orange). Both rise and fall times are 1~ms.  (b)-(c) Detailed views of (a), focusing on the periods of increasing and decreasing current. The red lines in (a)-(c) illustrate the applied power over time (right axis).
}
\end{figure*}

We first ran time-dependent simulations with a power of $P$ = 1~W applied to the electrode. We recorded the time evolution of the average temperature increase across the volume, $\Delta T$. Both the rise and fall times of the current were set to 1~ms based on our experiments. In this simulation, we used $l$ = \SI{120}{\micro\meter}, $w$ = \SI{20}{\micro\meter}, and $\Lambda_\mathrm{Si}$ = \SI{140}{\watt\meter^{-1}\kelvin^{-1}}, where $\Lambda_\mathrm{Si}$ is the thermal conductivity of \ce{Si}. The results in Fig.~\ref{Fig:TimeDependent}(a) show that $\Delta T$ rapidly approaches approximately \SI{90}{\percent} of the maximum temperature in 1~ms. Furthermore, $\Delta T$ scales almost linearly with $P$ when using such slow rise and fall times as mentioned in the main text [Figs.~\ref{Fig:TimeDependent}(b) and \ref{Fig:TimeDependent}(c)]. Based on these findings, we used a steady-state model to calculate the effective thermal resistance. \par

\section{Effective thermal resistance of substrates}

\begin{figure*}[t!]
\centering
\includegraphics[width = 8.5cm]{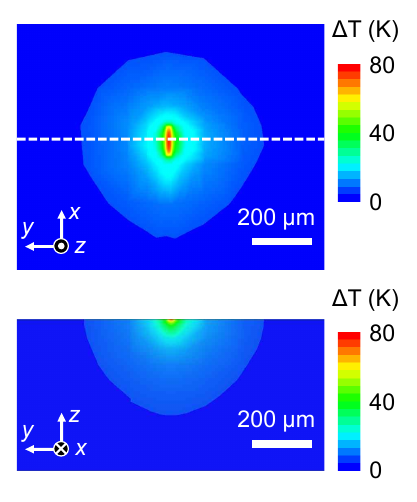}
\caption{\label{Fig:TempDistribution}
    Temperature distribution at steady state obtained from a simulation. A power of $P$ = 1~W is applied to the electrode. The length and width of the electrode are \SI{120}{\micro\meter} and \SI{20}{\micro\meter}, respectively.
}
\end{figure*}

We performed numerical simulations to determine the effective thermal resistance, $R_\mathrm{\theta, Si}$, of a bulk Si substrate in steady state. This calculation involved varying parameters: a diffusivity of the substrate, $D$, the width, $w$, and length, $l$, of an electrode. Figure~\ref{Fig:TempDistribution} shows a representative temperature distribution in the substrate when $l$ = \SI{120}{\micro\meter} and $w$ = \SI{20}{\micro\meter}. In this case, the temperature is largely distributed in a circular shape on the top surface of the substrate, which cannot be explained by the two-dimensional half-cylinder model~\cite{cahill_thermal_nodate}. \par

 Based on the two-dimensional analytical model, we assumed that

\begin{equation} \label{Eq:RthSi1}
\begin{split}
	R_\mathrm{\theta,Si} = \frac{\Delta T}{P} = \frac{\ln A}{\pi \Lambda_\mathrm{Si} l},
\end{split}
\end{equation}

with an unknown constant, $A$. $\Delta T$ represents the temperature increase due to Joule heating at steady state \par

\begin{figure*}[t!]
\centering
\includegraphics[width = 16cm]{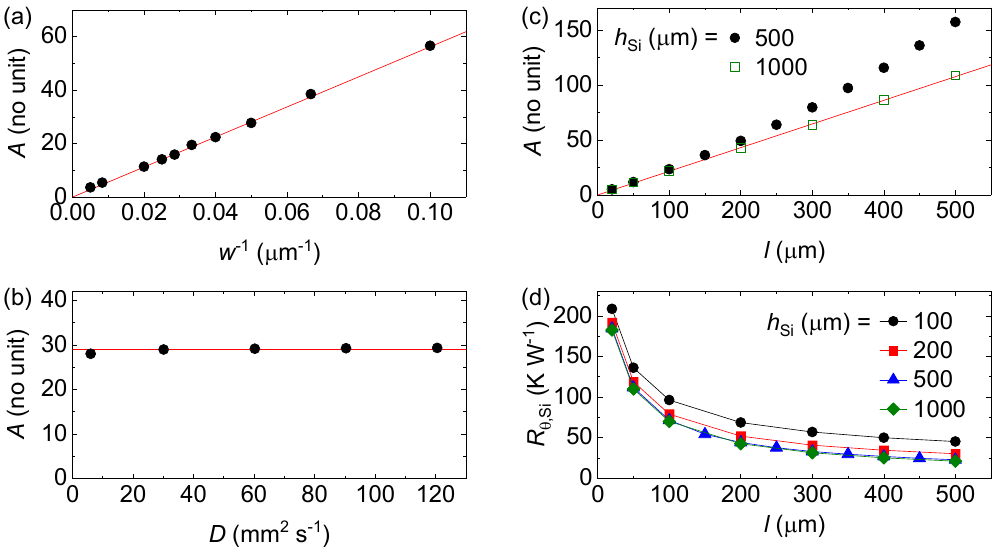}
\caption{\label{Fig:FEM}    
    (a)-(c) Dependence of the parameter $A$ in Eq.~\ref{Eq:RthSi1} on $w$, $D$, and $l$, from simulations. (d) $R_\mathrm{\theta, Si}$ calculated from simulations as a function of $l$ for different thicknesses of the Si substrate, $h_\mathrm{Si}$. In (a)-(d), $l$ = \SI{120}{\micro\meter}, \SI{20}{\micro\meter}, $D$ = \SI{84.4}{\milli\meter^2\second^{-1}}, and $h_\mathrm{Si}$ = \SI{500}{\micro\meter}, unless otherwise noted.
}
\end{figure*}

From the simulations, we obtained the variable $A$ in Eq.~\ref{Eq:RthSi1} for different $w$, $D$, and $l$. We used the following values: $l$ = \SI{120}{\micro\meter}, \SI{20}{\micro\meter}, $D$ = \SI{84.4}{\milli\meter^2\second^{-1}}, and $h_\mathrm{Si}$ = \SI{500}{\micro\meter}, unless specified otherwise. The results, shown in Figs.~\ref{Fig:FEM}(a) - (c), show that $A$ is inversely proportional to $w$ and nearly independent of $D$. $A$ is proportional to $l$, but it increases more sharply when $l \sim h_\mathrm{Si}$. The exceptional increase of $A$ is reduced when $h_\mathrm{Si} \gg l$ [green squares in Fig.~\ref{Fig:FEM}(c)]. Note that, as shown in Fig.~\ref{Fig:FEM}(d), $R_\theta$ also depends on $h_\mathrm{Si}$; it decreases with increasing $h_\mathrm{Si}$, but remains nearly constant when $h_\mathrm{Si} \geq l$. \par

From these simulations, we extrinsically obtained an analytical expression for the effective thermal resistance as

\begin{equation} \label{Eq:RthSi2}
\begin{split}
	R_\mathrm{\theta,Si} = \frac{\ln(\eta' l/w)}{\pi \Lambda_\mathrm{Si} l},
\end{split}
\end{equation}

where the constant $\eta' \approx 5$. From this result, we obtained Eq. 1 in the main text. \par

\section{Formation and rotation of octupole moments from a random state}
\begin{figure*}[t!]
\centering
\includegraphics[width = 8.5cm]{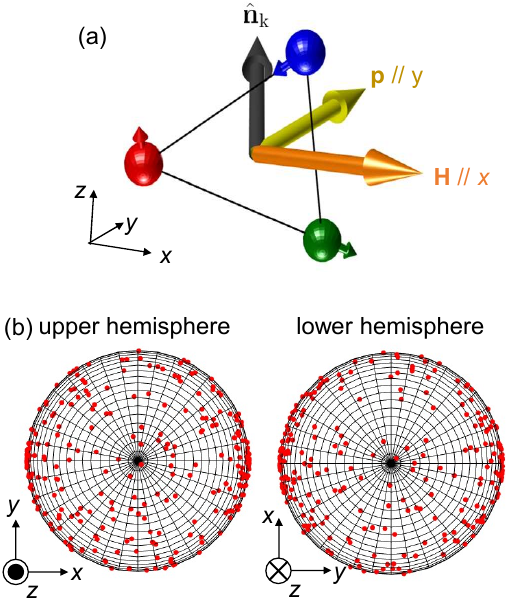}
\caption{\label{Fig:PlaneDistribution}
    (a) Schematic of the numerical calculation model. The spheres and small arrows show sub-lattice spins. The orange, yellow, and gray arrows represent a magnetic field, $\mathbf{H}$, the spin-orbit torque, $\mathbf{p}$, and the normal vector of a kagome plane, $\hat{\mathbf{n}}_\mathrm{k}$. The field and the spin-orbit torque are parallel to the $+x$- and $+y$-directions, respectively. (b) Distribution of $\hat{\mathbf{n}}_\mathrm{k}$ for numerical calculations. The left and right figures show an upper and a lower hemisphere, respectively.
}
\end{figure*}

We performed numerical calculations to simulate the dynamics of octupole formation and rotation in  a polycrystalline \ce{Mn3Sn} film. In these simulations, we maintained a magnetic field of 0.1~T in the $+x$-direction and fixed the spin-torque direction along the $y$-direction. To replicate a polycrystalline structure, we randomized the orientation of the kagome planes [see Figs.~\ref{Fig:PlaneDistribution}(a) and \ref{Fig:PlaneDistribution}(b)]. We created a thousand models, each with a unique set of initial spin states and kagome plane orientations, for each current density, $j_\mathrm{hm}$, in the heavy-metal layer. Figure.~\ref{Fig:PlaneDistribution}(b) shows the distribution of the normal vectors, $\mathbf{n}_\mathrm{k}$, of the kagome planes for $j_\mathrm{hm} = 6 \times 10^{11}$ \SI{}{\ampere\meter^{-2}}. \par

\begin{figure*}[t!]
\centering
\includegraphics[width = 16cm]{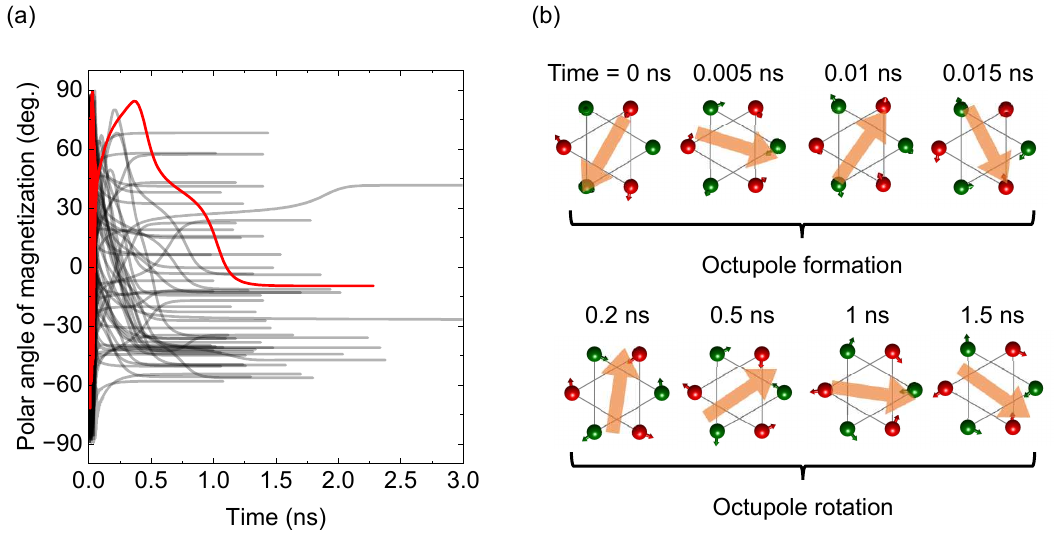}
\caption{\label{Fig:TimeDependentNumerical}
    Numerical simulations of the formation and rotation of magnetic octupole moments. (a) Time evolution of the magnetization angles during the formation of octupoles from random states and their subsequent rotation. In these simulations, octupoles are mostly formed within 0.1~ns. Here we used $j_\mathrm{hm} = +6 \times 10^{11}$ \SI{}{\ampere\meter^{-2}} and $\mu_0 H_x$ = +0.1 T. (b) Magnetic dipole moments of Mn sub-lattices during the formation (top rows) and rotation (bottom rows) processes in one of the simulations in (a) (red line). The orange arrows indicate the magnetization direction with in the kagome plane, which corresponds to the octupole orientation in the rotation process.
}
\end{figure*}

In our calculations, the octupole state, {\em i.e.}, the anti-chiral structure, is formed within 0.1~ns from a random state [Fig.~\ref{Fig:TimeDependentNumerical}(a)]. During this initial formation phase, the influence of the spin-orbit torque is minimal due to the predominance of strong exchange interactions. However, after the octupole state is established, the octupole moment begins to slowly rotate under the influence of the spin-orbit torque. Representative spin configurations during the formation and rotation phases of the octupole are shown in Fig.~\ref{Fig:TimeDependentNumerical}(c). \par

\section{References}
\bibliographystyle{apsrev4-2}
\bibliography{references}{}

%

\begin{figure*}[t!]
\centering
\includegraphics[width = 8.5cm]{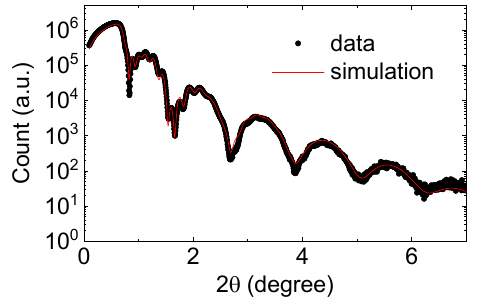}
\caption{\label{Fig:XRR}
    X-ray reflection spectrum of \ce{W/Mn3Sn} film. The black dots and the red line indicate the experimental data and the corresponding fit line, respectively. The thicknesses of \ce{W} and \ce{Mn3Sn} are about 7.1~nm and 34.4~nm, respectively. The surface roughness of \ce{W} and \ce{Mn3Sn} films are about 0.5~nm and 1~nm, respectively.
}
\end{figure*}

\begin{figure*}[t!]
\centering
\includegraphics[width = 8.5cm]{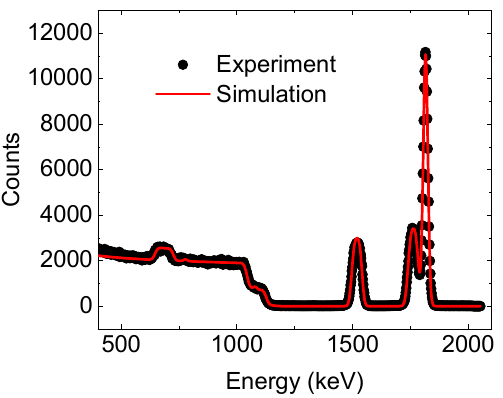}
\caption{\label{Fig:RBS}
    Rutherford backscattering spectrum of the \ce{W/Mn3Sn} film. The black dots and the red line indicate the experimental data and the fitted line, respectively. The atomic ratio between Mn and Sn in the film is Mn:Sn $\approx$ 77:23.
}
\end{figure*}

\begin{figure*}[t!]
\centering
\includegraphics[width = 8.5cm]{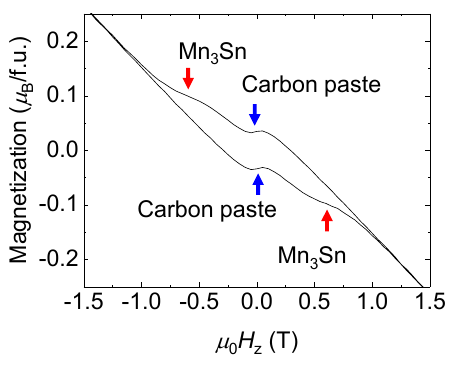}
\caption{\label{Fig:MPMS}
    Magnetization of the W/\ce{Mn3Sn} film as a function of the perpendicular magnetic field, $H_z$, at 300 K. The kinks indicated by red arrows show the magnetic responses of \ce{Mn3Sn}, whose magnetization is about 0.03 $\mu_\mathrm{B}$/f.u. and the coercivity is about 0.6~T. Blue arrows indicate an artifact due to residual carbon paste from the deposition process, which do not appear in the measurements of the anomalous Hall effect and the magneto-optical Kerr effect [see Figs. 1(c) and 2(b) in the main text].
}
\end{figure*}